\documentclass[12pt,preprint]{aastex}
\usepackage{natbib}
\usepackage{graphicx}
\shorttitle{Radio observation of the Andromeda galaxy}
\begin{document}
\title{Ruling out $\sim 100-300$ GeV thermal relic annihilating dark matter by radio observation of the Andromeda galaxy}
\author{Man Ho Chan$^1$, Lang Cui$^2$, Jun Liu$^{2,3}$, Chun Sing Leung$^4$}
\affil{$^1$Department of Science and Environmental Studies, The Education University of Hong Kong, Hong Kong, China \\
$^2$ Xinjiang Astronomical Observatory, Chinese Academy of Sciences \\
$^3$ Max-Planck-Institut f\"ur Radioastronomie, Auf dem H\"ugel 69, D-53121 Bonn, Germany \\
$^4$ Hong Kong Polytechnic University, Hong Kong, China}
\email{chanmh@eduhk.hk}

\begin{abstract}
In the past few years, some studies claimed that annihilating dark matter with mass $\sim 10-100$ GeV can explain the GeV gamma-ray excess in our Galaxy. However, recent analyses of the Fermi-LAT and radio observational data rule out the possibility of the thermal relic annihilating dark matter with mass $m \le 100$ GeV for some popular annihilation channels. By using the new observed radio data of the Andromeda galaxy, we rule out the existence of $\sim 100-300$ GeV thermal relic annihilating dark matter for ten annihilation channels. The lower limits of annihilating dark matter mass are improved to larger than 330 GeV for the most conservative case, which is a few times larger than the current best constraints. Moreover, these limits strongly disfavor the benchmark model of weakly interacting massive particle (WIMP) produced through the thermal freeze-out mechanism.
\end{abstract}


\section{Introduction}
Dark matter problem is one of the major problems in astrophysics. Many astrophysicists believe that there exists some unknown particles called dark matter particles which can account for the missing mass in galaxies and galaxy clusters. Many models suggest that dark matter particles can self annihilate to give photons, electrons, positrons and neutrinos. If dark matter particles are thermal relic particles (the simplest model), standard cosmology predicts that the thermal relic annihilation cross section is $\sigma v=2.2 \times 10^{-26}$ cm$^3$ s$^{-1}$ for dark matter mass $m \ge 10$ GeV \citep{Steigman}. The latest gamma-ray observations of Milky Way dwarf spheroidal satellite (MW dSphs) galaxies constrain the dark matter mass $m \ge 100$ GeV with the thermal relic annihilation cross section for the most popular $b\bar{b}$ quark and $\tau^+\tau^-$ annihilation channels (the Fermi-LAT limits) \citep{Ackermann,Albert}. Furthermore, by using gamma-ray data of two galaxy clusters (Fornax and A2877), \citet{Chan} also obtain $m \ge 100$ GeV for the $\tau^+\tau^-$ and $b\bar{b}$ channels. These limits are the current tightest constraints based on gamma-ray data. On the other hand, the latest analysis based on the positron detection also gives tight constraints for annihilation dark matter. The lower limits of dark matter mass for the $e^+e^-$ and $\mu^+\mu^-$ annihilation channel is $m \approx 100$ GeV \citep{Bergstrom,Cavasonza}. Therefore, the observations of gamma rays and positrons rule out the possibility of $m \le 100$ GeV annihilating dark matter for four popular annihilation channels ($e^+e^-$, $\mu^+\mu^-$, $\tau^+\tau^-$ and $b\bar{b}$). In particular, the lower mass limit obtained for the $b\bar{b}$ channel strongly challenge the recent dark matter interpretation of the gamma-ray excess in our Galaxy ($m \sim 30-80$ GeV) \citep{Daylan,Calore,Abazajian}. The new observational constraints also give tension to the dark matter models suggested that explain the GeV gamma-ray excess \citep{Zhu,Buckley,Marshall,Anchordoqui,Modak,Berlin,Agrawal,Boehm,McDermott}. Besides, although many earlier studies put stringent constraints on the possibility that the GeV gamma-ray excess is originating from millisecond pulsars \citep{Hooper,Calore2,Cholis2,Hooper2}, the latest observations from Fermi-LAT show that the properties of disk and bulge pulsar populations are consistent with the spatial profile and energy spectrum of the Galactic Center gamma-ray excess \citep{Ajello}. The debate of this issue is still on-going \citep{Bartels} and it can certainly give significant impact on the dark matter explanation for the gamma-ray excess.

Besides gamma-ray and positron detection, radio observation is another important way to constrain annihilating dark matter. If dark matter annihilates, the electrons and positrons produced by dark matter annihilation in a galaxy would emit strong synchrotron radiation due to the presence of magnetic field, which can be detected by a radio telescope. Previous studies of archival radio data of some galaxies (e.g. M31, M33, NGC2976) \citep{Egorov,Chan2,Chan3,Chan4} and galaxy clusters \citep{Colafrancesco,Storm2} can get more stringent constraints for annihilating dark matter. The lower limit of dark matter mass can reach $m \approx 200$ GeV for some annihilation channels. Nevertheless, \citet{Cholis} show that the effects of inverse Compton scattering and strong convective wind in our Galaxy are very important. The resulting radio constraints are therefore weaker than those derived from gamma-ray observations. In fact, that study considered the radio observations of a very small region ($r \le 0.14$ pc) in our Galactic Center so that these effects are extremely important. The strong wind and the strong inverse Compton scattering suppress the synchrotron radiation and significantly weaken the constraints for annihilating dark matter. However, these effects would be less significant and the constraints can be much tighter if the region of interest (ROI) considered is much larger (e.g. the entire galactic region). Generally speaking, choosing an appropriate target object, a large enough region of the target object and a right observing radio frequency can get very stringent constraints for annihilating dark matter. 

In this article, we report a new radio observation of the Andromeda galaxy (M31) by the NanShan 26-m Radio Telescope (NSRT) in Xinjiang Astronomical Observatory (XAO). We observe the radio wave emitted by the Andromeda galaxy. Based on the new observed data, the lower limits of dark matter mass for four popular annihilation channels can be improved to $\approx 350$ GeV, which is a factor of 2-3 larger than the current best constraints. These limits rule out the possibility of 100-300 GeV annihilating thermal relic dark matter, which largely narrow down the possible parameter space of the most popular candidate of dark matter, the weakly interacting massive particle (WIMP).

\section{Data collection}
The flux measurements of the Andromeda galaxy were carried out by using the Nanshan Radio Telescope (NSRT) on $24^{\rm th}$ August, 2017 (MJD 57990.33) and $24^{\rm th}$ October, 2017 (MJD 58049.98) at C-band frequency $\nu=4.6-5.0$ GHz. The observations were performed with cross-scans, where the antenna beam pattern was driven orthogonally over the source position, stacking 32 sub-scans for reaching the desired sensitivity. Frequent observation of non-variable calibrators allowed monitoring of the antenna gain variations with elevation and time, thus improving the subsequent flux density calibration. The data calibration was done in the well established standard manner that enables high precision flux density determination \citep{Kraus}.

As the first step of data calibration, for each scan, the sub-scans in each scanning direction are folded and averaged. The folding is performed through a precise positional alignment, in turn allowed by the very reliable antenna pointing system. Then a Gaussian profile is fitted to each averaged sub-scan. Due to the extended structure of the Andromeda galaxy comparing to the antenna beam size, the integral, rather than the amplitude of the Gaussian, is applied as a measure of the flux density. The integral of the averaged sub-scans in one cross-scan is then averaged. Subsequently we correct the measurements for the elevation-dependent sensitivity of the antenna and systematic time-dependent effects, using calibrators (e.g. 3C48) close to the targets. Finally the measured intensity is converted to the absolute flux density with a scaling factor determined by utilizing the frequently observed primary calibrators 3C286, 3C48 and NGC7027 \citep{Baars,Ott}. 

The final measurement errors are derived from the formal statistical errors and error propagation in each step of data calibration, which is mostly dependent on the root-mean-square of the measurement and the final calibration precision for calibrators. Since the angular separation between the Andromeda galaxy and the calibration source 3C48 is small, the precision of calibration is very good. Also, since we are focussing on the dark matter annihilation signal (mainly in the central region of the galaxy) but not the fine structure due to the baryonic processes, the calibration method is valid.

For the possible systematic uncertainties, the following factors have been considered. 1) In practice, the pointing error seldomly affects the flux density by more than 2-3\% \citep{Marchili,Marchili2}. 2) The gain induced variation is of the order 3-4\% \citep{Marchili,Marchili2}. 3) The flux dispersion of the calibrators is in the range between 0.2-0.7\% \citep{Marchili}. 4) The time-dependent variation is smaller than 2\% because of the good weather for data collection. Therefore, considering all of the above factors, the overall systematic uncertainties are not very significant.

The data obtained on two different days are consistent with each other. For NSRT, the beam profile can be approximately given by a Gaussian function with half-power-beam-width $\sim 550$ arc second. By choosing the data set from the most reliable measurement (i.e. with the best Gaussian width, elevation dependence and signal-to-noise ratio), we find that the integrated diffuse radio flux emitted by the galaxy is quite small: $S=82.9 \pm 11.6$ mJy (1 Jy = $10^{-26}$ W m$^{-2}$ s$^{-1}$ Hz$^{-1}$). The 2$\sigma$ upper limit of the radio energy flux is $5.1 \times 10^{-15}$ erg cm$^{-2}$ s$^{-1}$. The Gaussian fit of the radio flux of the Andromeda galaxy is shown in Fig.~1. Note that there are some side-band amplitudes on the Gaussian signal. This is expected because of the nature of the antenna beam pattern (side lobes) and the ring structure of the galaxy. However, it does not affect the overall Gaussian signal fitted significantly.

Note that the flux obtained $S=82.9 \pm 11.6$ mJy is much smaller than the one obtained in previous studies $S \sim 1800$ mJy \citep{Berkhuijsen2}. In fact, these two fluxes cannot be compared directly as these values are obtained by different observational methods. The observations in \citet{Berkhuijsen2} performed high-resolution survey which obtained every detail and fine structure in radio emission, including most of the strong point-source emissions, small-scale baryonic processes and large-scale emission. In our observations, we use cross-scan which is not a high-resolution survey. The method of cross-scans only observes the regions overlapped with the orthogonal axes, but not the whole galaxy. Those point-source emissions and small-scale baryonic processes outside the orthogonal axes would be omitted. In fact, the radio signal emitted by the Andromeda galaxy can be simply divided into two different contributions: 1) dark matter annihilation and 2) baryonic processes. The radio signal of dark matter annihilation is large-scale, smooth, continuous, spherically symmetric and centralized. Therefore, this radio flux contribution can be treated as a `global background radio flux' of the galaxy. Due to the spherically symmetric and centralized distribution, the overall dark matter contribution can be mapped or calculated by the signal obtained using cross-scans. For the radio signal due to baryonic processes, the signal is mainly discrete, localized, small-scale and dependent on different locations \citep{Berkhuijsen2}. Certainly, some large-scale emission (diffuse cosmic rays) and small-scale emission due to baryonic processes would overlap with the orthogonal axes and these contributions would be detected by cross-scans. However, the overall baryonic contributions detected would be significantly underestimated because these emissions are asymmetric and mainly in the ring structure. Generally speaking, the radio fluxes due to baryonic processes are very strong in small localized regions and they are much stronger than the `global background radio flux' due to dark matter annihilation. Although using cross-scan would miss most of the localized baryonic processes, the smooth global background signal due to dark matter annihilation (and some of the baryonic contributions) could be completely mapped by the cross-scan observations. In this study, our aim is to constrain the radio contribution due to dark matter annihilation only. It is not a problem even if we have missed all of the baryonic contributions. Therefore, using cross-scan may be a good method to eliminate many unwanted baryonic contributions automatically. A larger ratio of the dark matter contribution in the observed flux would give a more stringent constraint for dark matter annihilation.

\section{Dark matter annihilation model}
To constrain dark matter parameters, we assume that all of the radio flux observed originates from the synchrotron radiation of the electron and positron pairs produced by dark matter annihilation (i.e. $S=S_{DM}$). Since the average magnetic field strength in the Andromeda galaxy is large ($B=5 \pm 1~\mu$G \citep{Fletcher}), the cooling rate of the electron and positron pairs is dominated by synchrotron cooling and inverse Compton scattering. For a 10 GeV electron, the diffusion and cooling timescales are $\tau_d \sim R^2/D_0 \sim 10^{17}$ s and $\tau_c \sim 1/b \sim 10^{15}$ s respectively, where $R \sim 20$ kpc and $D_0 \sim 10^{28}$ cm$^2$ \citep{Berkhuijsen}. As the magnetic field in the outskirt of the Andromeda galaxy is still strong ($>4~\mu$G for $r \sim 38$ kpc \citep{Granados}), the cooling timescale is much shorter than the diffusion timescale for most of the parts in the galaxy. Since $\tau_d \gg \tau_c$, the diffusion term in the diffusion equation can be neglected and the equilibrium energy spectrum of the electron and positron pairs is simply related to the injection spectrum from dark matter annihilation \citep{Storm,Egorov}:
\begin{equation}
\frac{dn_e}{dE}= \frac{(\sigma v) \rho_{DM}^2}{2m^2b_{\rm total}} \int_E^m \frac{dN_e}{dE'}dE',
\end{equation}
where $\rho_{DM}$ is the mass density profile of dark matter, $b_{\rm total}$ is the total cooling rate and $dN_e/dE'$ is the injected energy spectrum due to dark matter annihilation, which can be obtained numerically for various annihilation channels \citep{Cirelli}.

Since the cooling rate due to inverse Compton scattering ($b_{ICS}=0.94\times 10^{-16}U_{eV}E_{\rm GeV}^2$ GeV s$^{-1}$) has the same energy dependence of the cooling rate due to synchrotron cooling ($b_{syn}=0.025\times 10^{-16}B_{\mu}^2E_{\rm GeV}^2$ GeV s$^{-1}$), the total cooling rate can be characterized by a correction factor $C$ such that $b_{\rm total}=b_{ICS}+b_{syn}=(1+C)b_{syn}$, where $C=38U_{eV}B_{\mu}^{-2}$, $U_{eV}$ is the radiation energy density in eV cm$^{-3}$, $E_{\rm GeV}$ is the energy of the electrons or positrons in GeV and $B_{\mu}=B/{\rm 1~\mu G}$. For a typical galaxy, $U_{eV} \approx 0.6$ eV cm$^{-3}$ \citep{Longair}. By taking this typical value for the Andromeda galaxy, we get $C \approx 0.9$, which means the inverse Compton scattering contributes about 50\% of the total cooling rate. Note that the value of $U_{eV}$ would be larger near the center of the Andromeda galaxy, which gives a larger cooling rate for inverse Compton scattering. Nevertheless, the magnetic field strength near the center is also much larger ($B \sim 15-19~\mu$G) \citep{Giebubel}. Following the result in \citet{Egorov}, the contribution of inverse Compton scattering is somewhat less than 47\% of the total cooling rate near the center of the Andromeda galaxy (i.e. $C \le 0.9$). On the other hand, the luminosity profile of the Andromeda galaxy is decreasing in the outskirt region \citep{Courteau}. Since the radiation density is proportional to the luminosity density, the radiation density is also decreasing in the outskirt region. As the magnetic field strength is close to constant in the outskirt region (see the discussion below), the inverse Compton scattering cooling rate is smaller than the synchrotron cooling rate. Therefore, overall speaking, assuming the flat profile of $U_{eV}=0.6$ eV cm$^{-3}$ in our analysis would slightly underestimate the synchrotron cooling rate near the center and the outskirt region. Hence it gives a more conservative result for constraining dark matter annihilation. The other processes such as Bremsstrahlung and Coulomb cooling are negligible because the baryonic number density is very low $\sim 0.1$ cm$^{-3}$ and the energy of electrons and positrons produced by dark matter annihilation is very high ($E \sim 10$ GeV). Also, since the supermassive black hole is inactive and the star formation rate is low, no strong convective outflow has been reported for the Andromeda galaxy \citep{Li}. \citet{Haud} shows that the greatest outflow velocity is about $v \sim 60$ km/s. The convection timescale is $\tau_o \sim R/v>10^{16}$ s, which is much longer than the synchrotron timescale $\tau_c \sim 10^{15}$ s. Therefore, the convective outflow is insignificant.

The power for synchrotron emission (with energy $E$ and frequency $\nu$) takes the following form \citep{Profumo}:
\begin{equation}
P_{DM}(E,\nu,\vec{r})=\frac{\sqrt{3}e^3}{m_ec^2}B(\vec{r})F(\nu/\nu_c),
\end{equation}
where $\nu_c$ is the critical synchrotron frequency and $F$ is the synchrotron kernel function. Since the radio emissivity is mainly determined by the peak radio frequency (monochromatic approximation), by combining the electron spectrum in Eq.~(1) with Eq.~(2), the total synchrotron radiation flux density of the electron and positron pairs produced by dark matter annihilation at frequency $\nu$ can be given by \citep{Bertone,Profumo}:
\begin{equation}
S_{DM} \approx \frac{1}{4 \pi \nu D^2} \left[ \frac{9 \sqrt{3} (\sigma v)}{2m^2(1+C)} E(\nu)Y(\nu,m) \int \rho_{DM}^2dV \right],
\end{equation}
where $D=785 \pm 25$ kpc is the distance to the Andromeda galaxy, $E(\nu)=13.6(\nu/{\rm GHz})^{1/2}(B/{\rm \mu G})^{-1/2}$ GeV, and $Y(\nu,m)=\int_{E(\nu)}^m(dN_e/dE')dE'$. Recent studies show that the Navarro-Frenk-White (NFW) density profile \citep{Navarro} can give a very good agreement with the dark matter density of the Andromeda galaxy \citep{Sofue}. Also, the magnetic field of the Andromeda galaxy is quite ordered and uniform, except the very inner region \citep{Giebubel}. We take the average magnetic field strength $B=5 \pm 1$ $\mu$G in our analysis \citep{Fletcher}. Although this value is constrained within $r=14$ kpc, many studies indicate that the magnetic field is almost constant even in the outskirt of the Andromeda galaxy ($r \sim 38$ kpc) \citep{Han,Granados}. The fitted value for $r \le 38$ kpc is $B=4.7^{+0.6}_{-0.7}$ $\mu$G \citep{Granados}, which is fully consistent with the value we used. Therefore, the magnetic field strength used can be justified for our ROI $\sim$ 40 kpc. Some recent studies take the sum of a constant part and an exponential part to model the magnetic field profile so that the magnetic field rises in the central region \citep{Egorov,McDaniel}. Generally speaking, a larger $B$ would give a larger value of $S_{DM}$ (except for $e^+e^-$ and $\mu^+\mu^-$ channels). However, the central magnetic field is hard to be determined \citep{Egorov}. The variation can be as large as a factor of 10. Therefore, in order to get the most conservative constraints, we take a constant profile (the average magnetic field strength) to model the magnetic field. Fortunately, since the dependence of $B$ is small, the value of $S_{DM}$ calculated would not be underestimated very much. By putting various parameters to Eq.~(3), we can get the predicted radio flux for dark matter annihilation $S_{DM}$ as a function of $m$ and $\sigma v$ for each of the annihilation channels.

Taking the central frequency $\nu=4.85$ GHz and the average magnetic field $B=5~ \mu$G, the peak energy is $E=13.4$ GeV. If we take the thermal relic annihilation cross section $\sigma v=2.2 \times 10^{-26}$ cm$^3$ s$^{-1}$, Eq.~(3) can be simplified to 
\begin{equation}
S_{DM}=S_0 \left( \frac{m}{\rm 100~ GeV} \right)^{-2}Y(\nu,m),
\end{equation}
where $S_0$ is a constant which depends on the dark matter density profile only. Here, we consider two density profiles, the NFW profile and the Burkert profile \citep{Burkert}, to calculate $S_0$. By using the parameters of the NFW profile \citep{Sofue} and the Burkert profile \citep{Tamm} respectively, we get $S_0=1.85^{+0.03}_{-0.10} \times 10^3$ mJy (NFW) and $S_0=1.66^{+0.02}_{-0.10} \times 10^3$ mJy (Burkert). In particular, the Burkert profile can give the most conservative limits for dark matter mass. By setting $S=S_{DM}$, the lower limits of dark matter mass with the thermal relic annihilation cross section for different annihilation channels can be obtained. Furthermore, by treating the annihilation cross section as a free parameter, we can get the upper limit of $\sigma v$ versus $m$ for each annihilation channel.

In Fig.~2, we show the $2\sigma$ upper limits of $\sigma v$ for six popular annihilation channels (using the Burkert profile). The resultant upper limits are much tighter than the Fermi-LAT limits \citep{Ackermann,Albert}. If dark matter particles are thermal relic particles, the lower limits of dark matter mass are improved to 350-540 GeV, which is a factor of 2-3 larger than the current best radio limits \citep{Chan2,Chan3,Chan4} and a factor of 4-100 larger than the current best Fermi-LAT gamma-ray limits \citep{Ackermann,Albert} (see Fig.~2 and Table 1). In particular, the lower limit of the most popular channel $b\bar{b}$ is improved significantly to 410 GeV. We also extend our calculations to other four less popular annihilation channels (see Fig.~3). The resultant lower limits for the thermal relic annihilation cross section are larger than 330 GeV (see Table 1).

Recent simulations show that substructures of dark matter halo can boost the annihilation signal to a few times larger for a galaxy \citep{Moline}. This effect can be characterized by a boost factor $B_M$. It can be calculated by an empirical formula which depends on the virial mass of a galaxy $M$ \citep{Moline}. By taking the virial mass of the Andromeda galaxy $M=(1.83 \pm 0.39) \times 10^{12}M_{\odot}$ \citep{Sofue} and using the most conservative model ($\alpha=1.9$) in \citet{Moline}, the boost factor due to substructure contribution is $B_M=5.28 \pm 0.06$. Using this boost factor and the parameters of the NFW density profile of the Andromeda galaxy in \citet{Sofue}, we get $S_0 \sim 10^4$ mJy. The resultant lower limits of dark matter mass can be improved to larger than 1 TeV. However, the boost factor model depends on a number of uncertainties, such as the slope parameter $\alpha$, tidal effect and the dark matter density profile used. To get the most conservative results, we neglect the effect of boost factor and just present its effect in Fig.~2 for reference only.

\begin{figure}
\vskip 10mm
 \includegraphics[width=120mm]{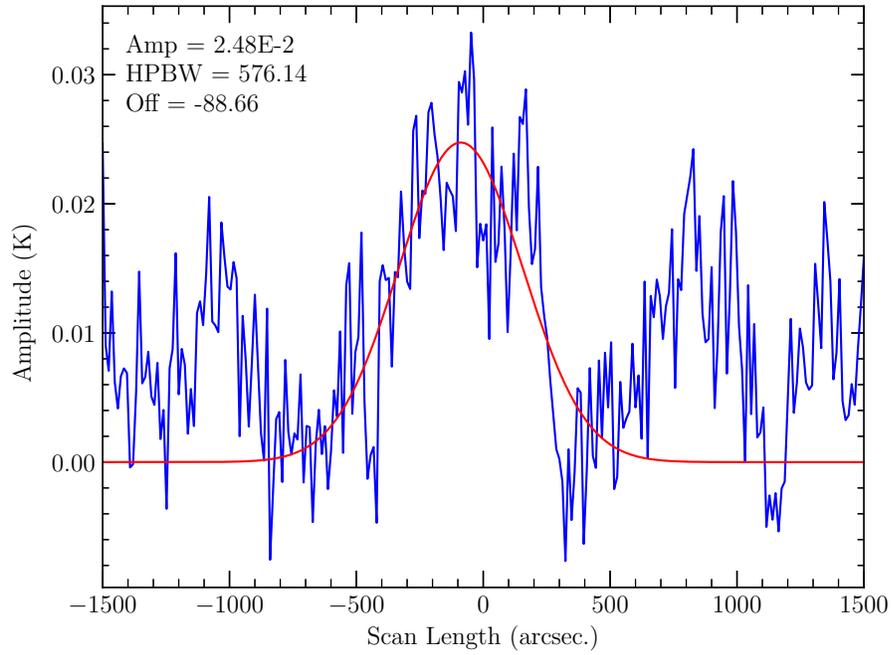}
 \caption{The Gaussian fit (red line) for the radio flux of the Andromeda galaxy.}
\vskip 10mm
\end{figure}

\begin{figure}
\vskip 10mm
 \includegraphics[width=120mm]{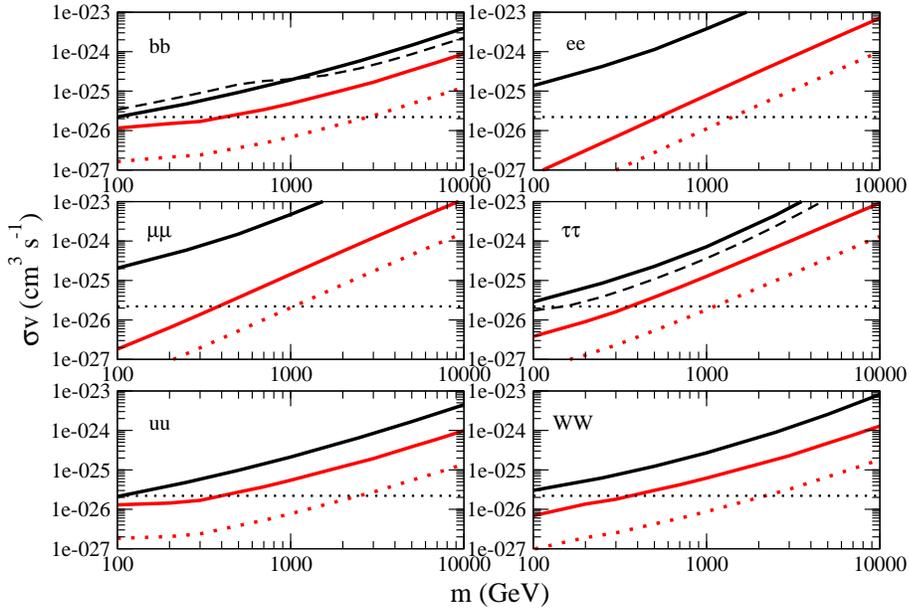}
 \caption{The $2\sigma$ upper limits of annihilation cross sections for six popular channels (red solid lines: Burkert profile without substructure contribution; red dotted lines: NFW profile with substructure contribution). The black solid and dashed lines are the $2\sigma$ Fermi-LAT upper limits \citep{Ackermann,Albert}. The black dotted lines indicate the thermal relic cross section \citep{Steigman}.}
\vskip 10mm
\end{figure}

\begin{figure}
\vskip 10mm
 \includegraphics[width=120mm]{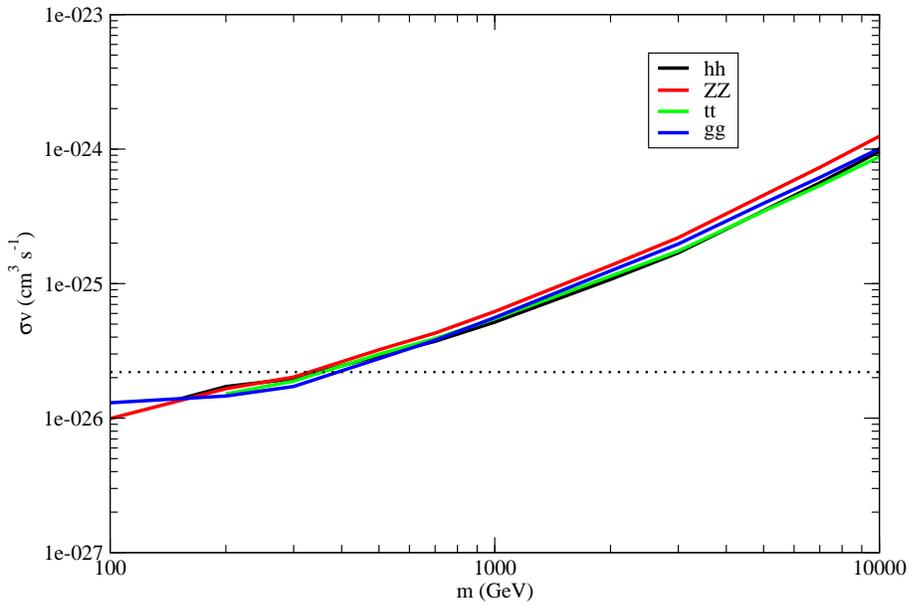}
 \caption{The $2\sigma$ upper limits of annihilation cross sections for four other channels (Burkert profile without substructure contribution). The dotted line indicates the thermal relic cross section.}
\vskip 10mm
\end{figure}

\begin{table}
\caption{The lower limits of dark matter mass (in GeV) for the thermal relic annihilation cross section. Columns 2 and 3 are the limits obtained in this study using the NFW and Burkert profiles respectively. The columns 4 and 5 are the current best radio limits \citep{Chan2,Chan3,Chan4} and the Fermi-LAT gamma-ray limits from the data of MW dSphs galaxies respectively \citep{Ackermann,Albert}.}
 \label{table1}
 \begin{tabular}{@{}lcccc}
  \hline
  Channel~~ &  NFW & Burkert & Current best radio & Fermi-LAT \\
  \hline
  $b\bar{b}$ & 470 & 410 & 100 & 100 \\
  $e^+e^-$ & 580 & 540 & 200 & 15 \\
  $\mu^+\mu^-$ & 390 & 370 & 130 & 3 \\
  $\tau^+\tau^-$ & 380 & 350 & 110 & 70 \\
  $u\bar{u}$ & 450 & 400 & 90 & 100 \\
  $W^+W^-$ & 410 & 370 & 90 & -- \\
  $gg$ & 440 & 400 & -- & -- \\
  $t\bar{t}$ & 400 & 360 & -- & -- \\
  $ZZ$ & 370 & 330 & -- & -- \\
  $hh$ & 400 & 350 & -- & -- \\
  \hline
 \end{tabular}
\end{table}

\section{Discussion}
In fact, the study in \citet{Egorov} has performed a detailed and systematic analysis based on the archival radio data of the Andromeda galaxy. However, the ROI in that study only focusses on a very small region ($\sim 1$ kpc) at the center. Since baryonic components dominate the central region of the galaxy (more than 97\% of mass is baryonic mass within 1.5 kpc \citep{Sofue}), the observed radio flux would be dominated by the baryonic processes. This would overestimate the upper limit of $S_{DM}$. Besides, the dark matter parameters used in that study are based on the old data of virial mass in \citet{Corbelli}. In our study, we observe the radio data for the whole Andromeda galaxy so that the overall expected dark matter radio contribution would be more significant (more than 61\% is dark matter within 40 kpc). We also use the updated dark matter parameters (with smaller uncertainties) obtained in \citet{Sofue}. By using the updated dark matter parameters and the larger ROI, the resulting limits of annihilation cross sections can be improved by a factor of about 2.7. Furthermore, using the new dark matter parameters and a larger ROI can reduce the uncertainties of the analysis. The uncertainties for the new scale radius and scale density are 6\% and 10\% respectively \citep{Sofue} while the corresponding uncertainties are 17\% and 25\% for the one used in \citet{Egorov}. The total uncertainty due to these dark matter parameter is reduced from about 110\% to 40\%. Also, the larger ROI can minimize the uncertainties because it can be characterized by the constant magnetic field strength with only 20\% uncertainty. However, for the ROI used in \citet{Egorov}, the magnetic field strength is not a constant and it is highly determined by the functional form and the central magnetic field strength ($\sim 15-300~\mu$G). The overall uncertainty can differ by a factor of 4. Consequently, the resultant conservative limits of dark matter mass obtained in \citet{Egorov} are less stringent (3$\sigma$ limit: $\sim 20$ GeV).

Our results improve the current lower limits of annihilating dark matter mass to 350-540 GeV for the most popular channels ($e^+e^-$, $\mu^+\mu^-$, $\tau^+\tau^-$ and $b\bar{b}$), which strongly challenge the dark matter interpretation of GeV gamma-ray excess ($m \sim 30-80$ GeV) \citep{Daylan,Calore,Abazajian} and the positron excess ($m \sim 50-100$ GeV) \citep{Mauro} for the thermal relic dark matter. Besides, the antiproton data from AMS-02 show a significant indication of dark matter signal near $m \sim 60-100$ GeV with the thermal relic cross section \citep{Cuoco,Cui}. However, our results also strongly rule out this suggestion. Furthermore, recent findings from the `Dark Matter Particle Explorer (DAMPE)' mission suggest that some excess electrons appear at $E \approx 1.4$ TeV in the observed energy spectrum \citep{Ambrosi}. Some suggest that this excess can be explained by dark matter annihilation with $m \sim 1.2-1.9$ TeV via $e^+e^-$ channel (with $\sigma v \sim 10^{-23}$ cm$^3$ s$^{-1}$) \citep{Jin,Niu}. Nevertheless, our results rule out this possibility because the lower limit of $m$ is $\sim 4$ TeV for $e^+e^-$ channel with $\sigma v \ge 10^{-24}$ cm$^3$ s$^{-1}$. More observations can further verify or falsify this proposal.

In this analysis, the major systematic uncertainties are the magnetic field strength profile and the dark matter density profile. We take a constant profile (the average value) to model the magnetic field so that we can avoid using the highly uncertain central magnetic field strength. As discussed above, a larger value of $B$ would give a larger value of $S_{DM}$ for most of the annihilation channels. Since the central magnetic field is larger than the average value, using the constant profile would somewhat underestimate the value of $S_{DM}$ and hence give conservative limits for $m$. Also, the magnetic field scale radius is about 1.5 kpc \citep{McDaniel}, which is much smaller than our ROI ($\sim 40$ kpc). Using the average magnetic field value can give a very good approximation and avoid the systematic uncertainty of the magnetic field profile assumed. However, if the magnetic field strength is much smaller in the outskirt region, the diffusion process in that region would be important. As a result, the limits obtained would be less stringent. For the dark matter profile, we use the NFW and Burkert profiles to model the dark matter distribution. In particular, the latter one would give the most conservative limits for $m$ because the profile is cored at the center. Therefore, our analysis gives the most conservative limits of $m$ for 10 different annihilation channels.

In many galaxies, the integrated diffuse radio energy flux is quite large. For example, in dwarf galaxy NGC2976, the total integrated diffuse radio power is $2.25 \times 10^{36}$ erg s$^{-1}$ at $\nu=4.85$ GHz \citep{Drzazga}. For the Andromeda galaxy, the total integrated diffuse radio power (including all baryonic processes) observed is $6.66 \times 10^{36}$ erg s$^{-1}$, which is about three times larger than that of the dwarf galaxy NGC2976. However, the size of the Andromeda galaxy is more than 10 times larger than that of the dwarf galaxy NGC2976. Therefore, the radio power density for the Andromeda galaxy is relatively small and it is the reason why its diffuse radio data can provide very tight constraints for annihilating dark matter. It suggests that the Andromeda galaxy is a very good target object in constraining annihilating dark matter.

Recent direct-detection experiments give negative results on dark matter search. Either the interaction cross section between dark matter and normal matter is very small or dark matter mass is larger than our expected value \citep{Aprile,Cooley}. The benchmark model of WIMP dark matter in which the observationally inferred dark matter abundance is obtained through the thermal freeze-out mechanism predicts $m \approx 220$ GeV with $\sigma v \approx 10^{-26}$ cm$^3$ s$^{-1}$ \citep{Bertone2}. Besides, a reconstruction from direct detection and gamma ray observations suggests a benchmark point with $m=250$ GeV and $\sigma v \sim 10^{-26}$ cm$^3$ s$^{-1}$ via $b\bar{b}$ channel \citep{Roszkowski}. However, these models and suggestions are ruled out by our results. Therefore, the actual production mechanism of dark matter may be much more complicated (non-thermal production) or dark matter mass is larger than $\sim 500$ GeV (thermal production). In view of this, opening the `TeV window' for detection is crucial for searching dark matter signal in future. More observational data from radio observations, DAMPE mission and High Energy Stereoscopic System (H.E.S.S.) \citep{Abdallah} can help solve the dark matter problem in astrophysics.

\begin{acknowledgements}
The work described in this paper was partially supported by a grant from the Research Grants Council of the Hong Kong Special Administrative Region, China (Project No. EdUHK 28300518). L. Cui and J. Liu thank for the grants supported by the CAS `Light of West China' Program (Grant No. 2015-XBQN-B-01), the National Natural Science Foundation of China (NSFC, Grant No. 11503072 \& 11503071), the National Key R\& D Program of China (Grant No. 2018YFA0404602), the Xinjiang Key Laboratory of Radio Astrophysics (Grant No. 2016D03020). L. Cui also thanks for the support by the Youth Innovation Promotion Association of the Chinese Academy of Sciences (CAS). We thank for the observation time of NSRT operated by Xinjiang Astronomical Observatory (XAO).
\end{acknowledgements}

\end{document}